# Survey on Distributed Data Mining in P2P Networks

**Rekha Sunny T**
Department of Computer Science and Engineering
Rajagiri School of Engineering and Technology, Kochi, India

**Sabu M. Thampi**
Indian Institute of Information Technology and Management - Kerala
Technopark Campus, Trivandrum-695581, Kerala, India

**ABSTRACT**

The exponential increase of availability of digital data and the necessity to process it in business and scientific fields has literally forced upon us the need to analyze and mine useful knowledge from it. Traditionally data mining has used a data warehousing model of gathering all data into a central site, and then running an algorithm upon that data. Such a centralized approach is fundamentally inappropriate due to many reasons like huge amount of data, infeasibility to centralize data stored at multiple sites, bandwidth limitation and privacy concerns. To solve these problems, Distributed Data Mining (DDM) has emerged as a hot research area. Careful attention in the usage of distributed resources of data, computing, communication, and human factors in a near optimal fashion are paid by distributed data mining. DDM is gaining attention in peer-to-peer (P2P) systems which are emerging as a choice of solution for applications such as file sharing, collaborative movie and song scoring, electronic commerce, and surveillance using sensor networks. The main intension of this draft paper is to provide an overview of DDM and P2P Data Mining. The paper discusses the need for DDM, taxonomy of DDM architectures, various DDM approaches, DDM related works in P2P systems and issues and challenges in P2P data mining.

# 1. INTRODUCTION

As computing and communication over wired and wireless networks advanced, many pervasive distributed computing environments such as internet, intranets, LANs, adhoc wireless networks, and P2P networks have emerged. These environments often deal with different distributed sources of voluminous data, multiple computing nodes, and distributed user community. Apposite utilization of these distributed resources should be guaranteed in mining such environments. Also local data sources can be of restricted availability due to privacy and as a result data sets at different sites must be processed in a distributed fashion without collecting everything to a single central site. Traditional data mining approach is to download the relevant data to a centralized location and then perform the data mining operations. Many of the distributed, privacy-sensitive data mining applications cannot make use of this centralized approach.

Distributed Data Mining (DDM) explores techniques of how to apply data mining in a non-centralized way. DDM requires an architecture which is totally diverse from the one used in centralized approach. In a distributed environment, the architecture must facilitate to pay careful attention to distributed resources of data, computing, and communication and human-computer interaction [1]. P2P networks are gaining growing status in many distributed applications such as file-sharing, web caching, network storage, searching and indexing of relevant documents and P2P network-threat analysis. It enables a collection of nodes (peers) to share computer resources in a decentralized manner. Collectively the peers already store a huge amount of widely varying data collected from different sources. If this data, distributed over large number of peers, can be integrated, it represents a very valuable data repository that, upon mining, may give very exciting and useful results [36]. Hence DDM in this domain is gaining increasing attention for advanced data driven applications.

This draft paper provides an overview of various DDM approaches and related works on P2P Mining. The paper is organized as follows: Section 2 describes centralized data mining approach and the issues associated with centralized approach. Section 3 discusses distributed data mining and taxonomy of DDM architectures. Section 4 describes the issues and challenges of DDM. Section 5 gives an overview of P2P networks. Section 6 introduces P2P data mining, presents the motivation, and identifies issues and challenges of P2P data mining. Section 7 briefly describes the related works on P2P data mining. Finally, Section 8 concludes the paper.

## 2. DATAMINING-CENTRALISED APPROACH

Data mining is the overall process of discovering new patterns or building models from a given dataset [11]. The steps in the KDD process include data selection, data cleaning and preprocessing, data transformation and reduction, data-mining task and algorithm selection, and finally post-processing and interpretation of discovered knowledge [11, 12]. This KDD process is highly iterative and interactive. Data mining algorithms fall under three major categories: clustering, frequent item set mining and classification. In traditional approach, data mining in particular, clustering has been done outside of a database. This involves shipping the data (possibly from multiple sources) to a single destination where all the processing takes place in a single processing computer [1]. The architecture for centralized data mining is shown in figure 1. ADaM ,WEKA are some data mining suites that operate directly on a file structure which is outside of the database.

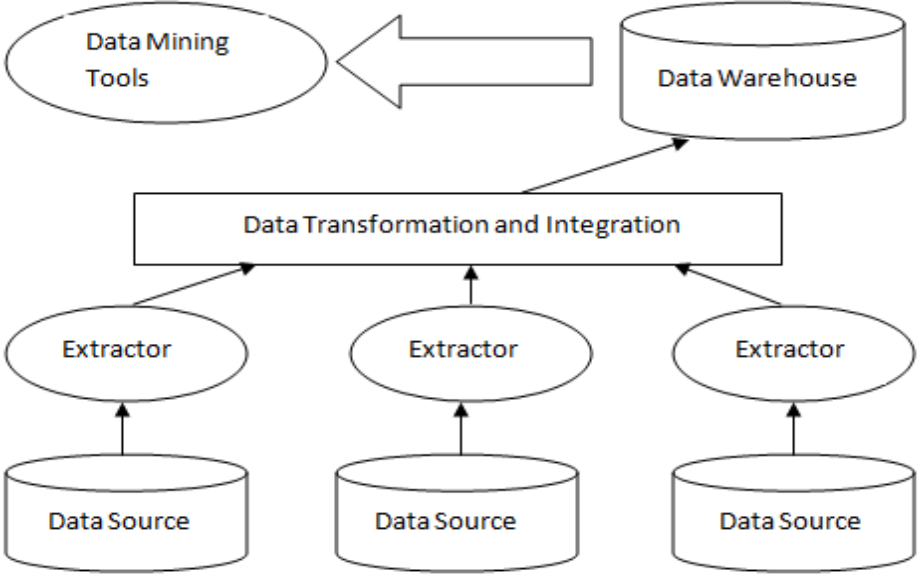

Fig 1: data mining –centralized approach   [1]

*Issues of Centralized approach*

Centralized data mining operate on a principle of gathering all data into a central site, then running an algorithm against that data (as shown in figure). Such an approach is fundamentally inappropriate for most of the distributed data mining applications leading to a need for distributed data mining. In fact, the long response time, lack of proper use of distributed resource, and the fundamental characteristic of centralized data mining algorithms do not work well in distributed environments.

There are several situations where need for distribution of data arises:

1. *Connectivity*: It is infeasible to transmit large quantities of data to a central site. As pointed out in [4], "Building a monolithic database, in order to perform non-distributed data mining, may be infeasible or simply impossible" in many applications. The cost of transferring large blocks of data may be exorbitant and result in very inefficient implementations. As an example, let us consider the World Wide Web which contains distributed data and computing resources. A growing amount of databases and data streams are currently made online, and changes occur very frequently upon this. Many applications exist that require regular monitoring of these diverse and distributed sources of data. In particular when the application involves a large number of data sites, a distributed approach to analyze this data is likely to be more scalable and practical. Hence, in this case we need data mining architectures that pay careful attention to the distribution of data, computing and communication, in order to access and use them in a near optimal fashion. Distributed Data Mining considers data mining in this broader context.

2. *Privacy of sources*: Organizations may be willing to share data mining results, but not data. The privacy issue is playing an important role in data mining applications. In many applications, mostly in security-related applications, data is confidential. Centralizing the distributed data sets is not tolerable in such cases. Hence data mining applications in such domains must analyze data in a distributed fashion without having to first download everything to a single site. Furthermore, these applications must pay cautious notice to the amount and type of information exposed to each site about the other sites' data. For example, consider a conglomerate of different banks collaborating for detecting frauds. In the centralized approach, all the data from every bank should be collected in a single location, to be processed by a data mining system. However in such a case a distributed data mining system should be the accepted choice, it is able to learn models from distributed data without exchanging the raw data between different repository and it allows detection of fraud by preserving the privacy of every bank's customer transaction data.

## 3. DISTRIBUTED DATAMINING

Distributed Data Mining is a framework to mine distributed data which operates on an architecture that is totally different from centralized approach. It cares the distributed sources of data, computing and communication.DDM architecture includes multiple sites each having independent computing power and storage capability. Each site performs local computation on its own and finally either a central site communicates with each distributed site to compute the global models or a peer-to-peer architecture is used. In the latter case, individual nodes perform most of the tasks by communicating with neighboring nodes by message passing over an asynchronous network.

The architecture for DDM is as shown in figure 2. From the figure it is clear that, in a distributed setting several local models are generated on different nodes and finally aggregated to form a global model which represents the mining result of the entire dataset.

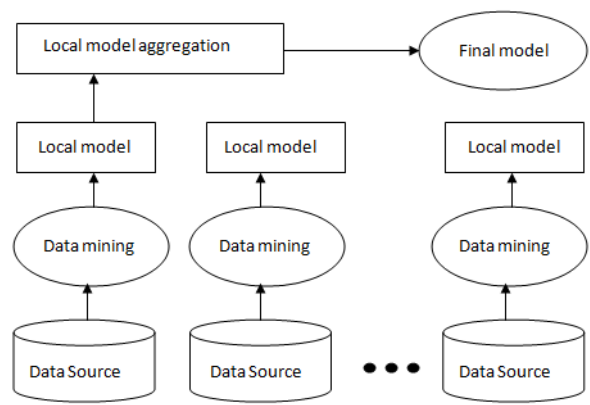

Fig 2: Data mining –DDM approach   [1]

## 3.1 Taxonomy of DDM Approaches and Related works

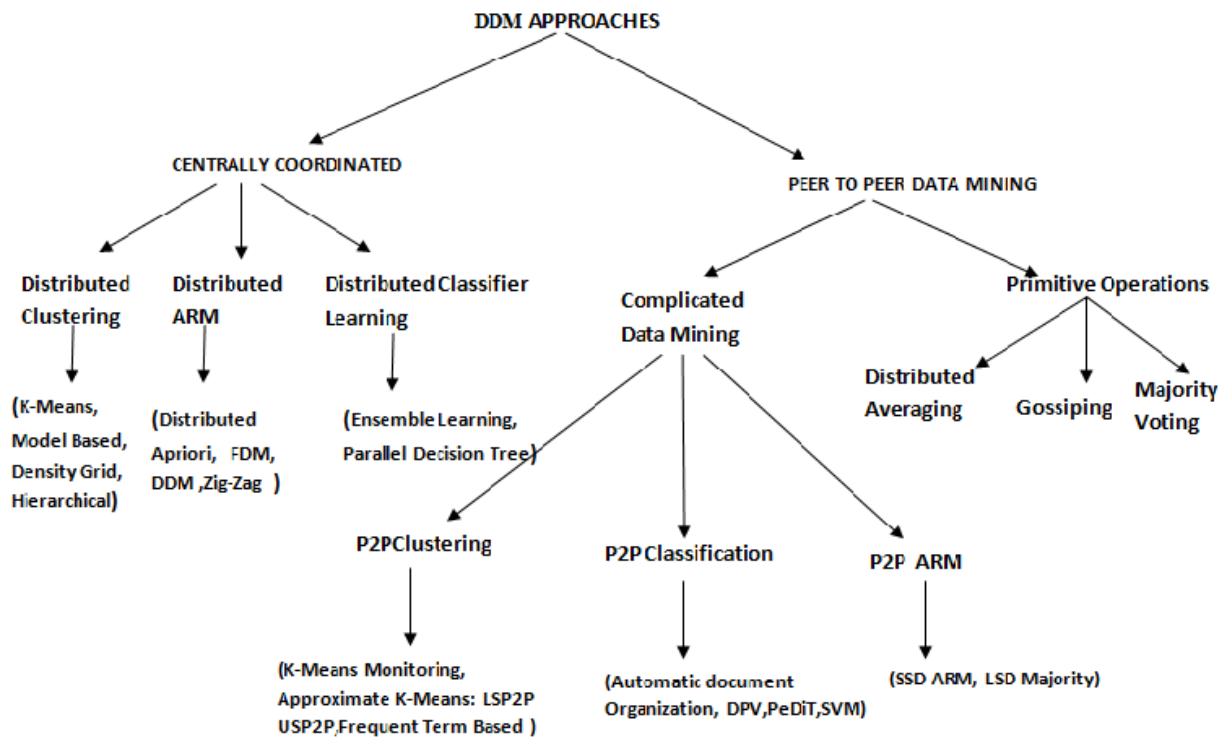

Distributed data mining comprises two types of architectures - centrally coordinate architecture and P2P [2]. In the first architecture, the entire data mining work is split into multiple workers and a central process coordinates the workers. However this approach suffers from the problem of single point of failure. Privacy issues and communications concerns are also associated with this. The second architecture, P2P data mining overcomes these problems where a large number of nodes are connected in an ad-hoc way. As communication is only with its neighbors, overhead is low and elegant handling of failure of single nodes.

Works on P2P data mining architecture is divided into primitive operations and complicated data analysis, described in detail in section 5. Major works in centrally co-ordinated architecture are related to distributed classifier learning, distributed clustering and distributed association rule mining.

**Distributed Clustering**

Database records are partitioned into clusters via clustering where elements of a cluster share a set of common properties that distinguish them from other clusters. The goal of clustering is to minimize inter-cluster similarity and maximize intra-cluster similarity. A clustering algorithm involves 3 steps in general. First step is to compute local models using local clustering algorithms. Next aggregate local models by a central node and finally either compute the global model or aggregated models are sent back to all the nodes to produce locally optimized clusters [2]. Some of the distributed clustering algorithms are K-means, model based, density grid and hierarchical.

**K-means**: Initially the k cluster centers among all the random points are randomly chosen. These k cluster centers are then sent to every local representative and local K-means clustering is performed. Each local machine then gathers the statistics about membership within its own clusters. Each of the statistics is then transmitted to a central controller to aggregate the models. As we are transferring the statistics rather than the entire data, data privacy is maintained. But this statistics need to be sent over and over again until convergence. This algorithm does not scale well and not assured to be a very quick process [4].

**Model Based**: This algorithm uses expectation maximization clustering [5] on the local level which is similar to K-means, except that decision on final clustering based on additional functions like Gaussian function. It is described in [6].Initially the local system processes its own individual pieces, by local EM clustering and each cluster is modeled as a sum of Gaussian functions. These functions are then transferred to a central coordinator which combines the functions to give global information about the probability density of the global picture. This information is then sent to each local source and they can make use of the new information, reevaluate the data if needed. This algorithm employs good measures for privacy and accuracy. This method suffers from a standard problem such that two large clusters with a small densely connected component can end up being in the same cluster even if they should not be so [4].

**Density Grid**: This algorithm makes use of the CLIQUE algorithm [9] with further improvements, suitable for distributed clustering. In [7] describes Density Based approaches for distributed clustering. Initially each of the attributes in the user query is scanned and then instead of setting some global setting for the size of each grid square they decide on it dynamically based on the statistics [4]. Here clusters are represented as a filling of a grid and due to the dynamic gridding, in heavy density areas granularity is finer and in areas of lower population granularity is coarse. This algorithm yields strong clusters. However, it is based on the assumption that data is centralized in a single repository from which it is distributed to other nodes or processors.

**Hierarchy**: [8] presents hierarchical algorithm which is similar to density gridding type approach. The general idea behind this algorithm is to start with a set of distinct points, each forming its own cluster and continue recursively merging two closer clusters into one until all points belong to a single cluster. Thus in parallel hierarchical algorithms, Dendograms are used to create clusters and a tight bound on the minimum and maximum distance between clusters is created. Here merging of clusters is based on the minimal statistics (specifying the minimum distances for merging) that is transmitted along with TID (object identifier). Reducibility property [8] is used in building the global model.

**Distributed Association Rule Mining (ARM)**

Research in distributed ARM is not very active, only few works such as distributed apriori, FDM, DDM and ZigZag come under this. In general, ARM uses the notion of frequent item sets to compute a result.

**Distributed Apriori Algorithms:** The simple distributed version of Apriori algorithm including count distribution and data distribution is presented in [10]. In count distribution, initially each node computes the candidates for the frequent k-item sets of its local data. The frequency values along with the candidates of nodes are transmitted among the nodes. Each node can then determine individually which item sets are frequent in the overall data by using the information transmitted and starts the next round to compute all candidate (k + 1)-item sets based on the set of frequent k-itemsets obtained. The communication cost of the Count Distribution algorithm is small since only candidates and their frequencies are sent. In data distribution all nodes compute disjoint sets of candidates. But because of big communication overhead performance of this algorithm is very poor.

**FDM**: Another version of distributed Apriori is Fast Distributed Mining of association rules is presented in [11. The difference from Count Distribution is in the message contents sent between the nodes. This algorithm belongs to shared-nothing architecture algorithms.

**DDM**: Distributed Decision Miner belongs to Apriori based algorithms assuming a shared-nothing architecture is presented in [12]. Initially local frequency counts are computed on each node, and then the nodes determine the set of globally frequent itemsets by performing a distributed decision protocol in each round. The local frequency counts of itemsets can be published by nodes. According to the protocol, if no node decides to publish its frequency count of a certain itemset, then it is known to the public hypothesis if this itemset is globally frequent. DDM is very communication-efficient due to the distributed decision protocol. DDM seems to be more scalable compared to other Apriori-based algorithms.

**ZigZag Algorithm**: This algorithm is based on the assumption that data is initially distributed on different sites in which each site first generates the local set of maximal frequent itemsets(MFI) [13]. The global set of MFI can then be computed. All frequent itemsets can be determined using the set MFI, infrequent candidates are not generated. A single scan of each local dataset is then sufficient to compute the frequency counts of the frequent itemsets. Communication in zigzag algorithm is fairly less than the traditional Apriori-based algorithms, since an exchange of the local models take place only after the local sets of MFI have been computed.

**Distributed Classifier Learning**

Most of distributed classifiers are based on ensemble learning which increase classification accuracy of predictive models. Basic idea is to produce base classifiers from homogenous data sets and aggregate those using voting schemes. Ada Boost , arcing , stacking are some of the aggregation approaches.

**Meta Learning:** Meta learning is an ensemble classifier approach in which base classifiers at each site are generated first using a classifier learning algorithm. Then the base classifiers are collected to a central site where meta level data from a separate validation set are produced. Finally, the meta classifier is generated from the meta level data. Arbiter scheme and combiner are two common techniques for meta learning from the output of the base classifiers. Communication overhead is very low in meta learning.

**Collective Data Mining:** CDM is proposed by Kargupta and his colleagues for predictive data modeling. CDM obtains local building blocks that directly constitute the global model [1].The first step in CDM is to generate approximate orthonormal basis coefficients at each local site. Then move an appropriately chosen sample of the data sets from each site to a single site and generate the approximate basis coefficients corresponding to non linear cross terms. Finally combine the local models and output. CDM framework is extended to Bayesian network learning. Other extension to CDM framework is distributed decision tree construction.

## 4. ISSUES AND CHALLENGES OF DDM

Even though DDM covers many of the issues associated with centralized approach, modern requirements in data mining inspired by emerging applications and the peculiarities of data sources leads to many challenges. The critical features of data sources determining such requirements are as follows:

In enterprise applications, data is distributed over many heterogeneous sources either tightly coupled or loosely coupled. Complexity of distributed data sources associated with a business line is very high due to mixing static and dynamic data, mixing multiple structures of data. Data integration and data matching are difficult to conduct. It is hard to store them in centralized storage and infeasible to process them in a centralized manner. Privacy is a major concern and hence availability of local data sources becomes restricted, thus prevents its centralized processing. In many cases, distributed data spread across global storage systems is often associated with time difference. Availability of data sources in a mobile environment depends on time. The infrastructure and architecture weaknesses of existing distributed data mining systems require more flexible, intelligent and scalable support. These peculiarities make DDM approach with central co-ordination infeasible and require the development of new approaches and technologies of data mining to identify patterns in distributed data. Distributed data mining (DDM), in particular, Peer-to-Peer (P2P) data mining is a solution to the above challenges [44].

## 5. OVERVIEW OF P2P NETWORKS

Peer-to-Peer (P2P) technology has emerged recently and become extremely popular due to its self- organization, flexibility, scalability, fault-resilience and robustness. Oram [43] gives a simple definition of peer-to-peer (P2P) as: "P2P is a class of applications that take advantage of resources storage, cycles, content, human presence available at the edges of the Internet. Because accessing these decentralized resources means operating in an environment of unstable connectivity and unpredictable IP addresses, peer-to-peer nodes must operate outside the DNS and have significant or total autonomy of central servers." Resource sharing, Decentralization, Dynamicity, Autonomy and self-organization are the principles that characterize P2P applications. In a P2P system resources are shared among all participants forming an overlay network. There is no need of centralized coordination and thus avoids a central point of failure. This feature makes P2P systems more scalable and robust. As there is no central coordinator peers are responsible for organizing themselves autonomously. Each of the peers is treated as a completely independent entity that has the flexibility to interact with its neighbors with a lot of flexibility. Volatility of the network connections is another key feature of P2P systems; peers operate outside the DNS, which is mainly characterized by its static nature, where nodes rarely change their topology. Dynamicity feature makes P2P applications fit perfectly with the Internet model. Peers can join and leave the P2P network at any time in a flexible manner without affecting or harming the entire system as a whole.

### *P2P Applications*

P2P technology is widely used in many different areas such as business, e-commerce, distributed computing, and communication. Napster, Gnutella, Morpheus, Freenet, and Kazaa are some of the most popular file-sharing applications. Napster [45] was used for sharing mp3 files between registered users. Chat rooms and querying Napster servers are some other services provided by Napster. Napster is considered as one of the first P2P applications, but it is not a pure P2P system. Napster maintains a central database that indexes all the participating users and the files that they are sharing. Many similar applications emerged after Napster with pure P2P features such as Gnutella, Freenet, and Kazaa which rely on distributed search mechanisms to address the limitations of Napster's central servers and to provide fault-tolerance and scalability.

Distributed P2P concepts have been also very suitable and successful in the area of distributed computing. They are especially useful for large computations that have a high degree of parallelism and a coarse correlation between their different components. A considerable speedup can be achieved if large calculations could be divided into completely independent and parallel parts that can run on separate nodes. Typical applications are SETI@home, Folding@home, and FightAids@home. P2P applications such as ICQ, AOL Instant Messaging and MSN Messenger are used for communication. Magi, Groove, Jabber are some of the P2P collaborative applications that require update mechanisms to provide consistency in multi-user environment.

*P2P Systems (structured and unstructured)*

P2P systems can be classified into unstructured and structured systems.

Structured systems are self-organizing, load balanced, and fault-tolerant systems. They have scalable guarantees on a number of hopes to answer a given query. They are all based on a distributed hash table interface. In structured systems, the overlay network assigns keys to data items and organizes its peers into a graph that maps each data key to a peer. This structured graph enables efficient discovery of data items using the given keys. However, in its simple form, this class of systems does not support complex queries and it is necessary to store a copy or a pointer to each data object (or value) at the peer responsible for the data object's key. Content Addressable Network (CAN) [47], Tapestry [48], Chord [49] and Viceroy [50] are some of the structured systems [46].

In unstructured systems, the overlay networks organize peers in a random graph in flat or hierarchical manners (e.g. Super-Peers layer) and use flooding or random walks or expanding-ring Time-To-Live (TTL) search, etc. on the graph to query content stored by overlay peers. Each peer visited will evaluate the query locally on its own content, and will support complex queries. Freenet [51], Gnutella [52], FastTrack [53], Kazaa [54] are some of the unstructured systems [46].

## 6. P2P DATA MINING

The server-less P2P networks are becoming very popular by the introduction of high speed network connectivity. Motivated by this, P2P mining has emerged recently from DDM. A huge amount of widely varying data from various sources is stored in these peers. Integrating these data and then mining will provide valuable results. For example, topic-wise document clustering in a P2P document repository help to organize the results returned by a search engine in response to a user's query. It should be ensured that data mining results by P2P mining are close to the data mining result by a centralization approach, provided there is no data movement from its original location. For this P2P algorithms must possess a unique set of characteristics like scalability, reliability, availability etc. There are a lot of application areas like MANET, Sensor networks which motivate P2P mining. For example, consider a simple profiling application launched via cellular phones that tries to automatically connect to MANET like network formed by different cellular phones in the vicinity and identify peers with similar interest to form a social network. A lightweight P2P classification algorithm will be very helpful in such an application [36].

### 6.1 Issues and Challenges in P2P Data Mining

Several issues are associated with mining on P2P networks. The issues like limited bandwidth, limited memory, limited CPU capacity and limited battery power are very crucial in the P2P computing environment since it reduces the efficiency of mining. Data dynamics is an important feature of P2P networks. While developing mining algorithms, changing nature of data in peers need to be considered.

High communication cost, non scalability are some of the challenges in P2P mining, efficient algorithms to overcome these to a great extent are discussed in [39, 37, 14, 15]. Privacy is a major concern in P2P networks, in order to address this issue privacy preserving data mining algorithms are developed which are discussed in [16, 17, 18]. Scalability, fault tolerance, decentralization, communication efficiency, anytimeness, asynchronism are some of the desirable operational characteristics for P2P mining algorithms.

- **Scalability**

The foremost characteristic a P2P data mining algorithm must possess is scalability. Each peer in a P2P network contains some data and scalability with respect to the size of the data is significant. Data mining algorithms for P2P networks should be either independent of the size of the system or at most dependent on log of the system size[36].

- **Communication Efficiency**

P2P data mining algorithms should be communication efficient. A data mining algorithm that is designed to analyze the large volumes of data stored in P2P systems must be able to work well with less exchange of data among the nodes. In effect the communication overhead should be minimized as much as possible while performing distributed data analysis.

- **Decentralization**

For tasks involving large volumes of data, centralized systems lead to huge data transfer and hence reduce the efficiency and quality of data mining. Hence P2P data mining algorithms should not use any centralized coordination.

- **Fault Tolerance**

P2P data mining algorithms must be able to work well even if some of the peers fail. In a P2P system, peers can leave or join the system at any time. So there are chances for loosing data and partial results as well as peer failures, hence algorithms should be able to recover from such situations.P2P data mining algorithms must be robust enough.

- **Anytimeness**

The data mining algorithms for P2P systems should be incremental. Since the data at the peers change very frequently, it is not suitable to design algorithms that need to begin from the scratch at every data change. In some applications the rate of data change will be more than the computation rate. Hence anytime algorithms that can report a partial ad hoc solution at any time is more suitable.

- **Asynchronism**

The number of nodes in P2P systems are very large, usually in the range of million. There are a lot of factors like limited bandwidth, connection latency that prevents successful synchronization between the entire P2P networks. Hence algorithms that require global synchronization are not suitable for P2P data mining.

## 7. RELATED WORKS ON P2P DATA MINING

P2P data mining has emerged recently from distributed data mining, motivated by the rapid growth of P2P networks. P2P data mining approaches have focused on developing algorithms for both primitive operations like average, count etc. and complicated data analysis and data mining.

Data mining which deals with complex approaches such as P2P association rule mining, P2P clustering, P2P Classification, P2P L2 threshold monitor and outlier detection in wireless sensor networks are grouped into complicated data mining. The primitive operations are extensively used in algorithms of complicated data mining.

### 7.1 Primitive Operations

Algorithms for primitive aggregates such as average, count, sum, max and min are developed by researchers which are described in [19, 20].In this approach, the value on each peer is computed only once. Other approach for processing sum and count that uses probabilistic counting is described in [21]. [22,23] presents gossip based aggregate computation which uses small messages and gossip-style local communication. In gossip, every peer sends its statistics to a random peer that can be used to compute a variety of aggregated statistics. Gossip algorithms provide probabilistic guarantee for the accuracy of their outcome. Even though gossip algorithms are simple and highly fault tolerant, for the computation of just one statistic they require hundreds of messages per peer. Approaches that rely on empirical accuracy results rather than guaranteed correctness are also developed for primitive operations [24].These approaches use epidemic protocols for computation. All these works laid a foundation for sophisticated data mining algorithms. These efficient primitives were used in developing efficient complex algorithms.

Further researches for primitive operations were based on developing local algorithms. Distributed averaging based on local algorithms are described in [25, 26]. In order to compute the average, [25] presents a graph Laplacian–based approach. Works on majority voting over P2P networks that requires no synchronization between the computing nodes are described in [27]. Algorithm for computing majority vote over P2P network is used for K-facility location which is described in [28]. All these researches for primitive operations led to the development of algorithms for complicated problems.

## 7.2 Complicated Data Mining

Researches for complicated problems include P2P association rule mining [27, 30], P2P clustering, P2P Classification [31, 32, 33, 34], P2P L2 threshold monitor [29] and outlier detection in wireless sensor networks [35]. In most of the works results are computed using information from their immediate neighbors that is in a decentralized manner.

### 7.2.1 P2P Association Rule Mining

In P2P systems data is distributed so widely and it needs to be processed without moving to anywhere. A lot of interesting knowledge can be discovered from this upon association rule mining. For example we can mine user preferences over the Kazza file sharing network. It is possible to discover knowledge like "people who are interested in songs of category A also look for category B songs ".This knowledge is like "customers who purchase diapers on Thursdays also buy beer". So in the P2P association rule mining algorithm the participating nodes must decide whether each itemset is frequent or not.

As this is a very young area, only few algorithms are developed for P2P association rule mining. [30] presents an algorithm for association rule mining using broadcast and global synchronization which seems to be suitable for small-scale distributed systems. Another work using majority voting protocol – LSD-Majority – which works well for large-scale distributed systems is described in [27].In that distributed association rule mining problem is reduced to distributed majority voting problem. The algorithm combines sequential association rule mining, executed locally at each node, with a majority voting protocol .The majority voting protocol discovers all of the association rules that exist in the combined database at each node.

### 7.2.2 P2P Classification

P2P classification helps to learn classification models from the distributed data sources with a provision of shared resources, such as bandwidth, storage space, and rich computational power. However the challenging issues such as scalability, peer dynamism, asynchronism and fault-tolerance limits the researches in P2P classification.

One of the works in P2P classification is on automatic document organization in a P2P environment which is presented in [31].It mainly focuses on accuracy issue, in which with respect to the user specified topic catalogue, web pages are gathered on each peer. A local predictive model is derived based on the training set, which is used to classify these web pages into specific topics. Finally a global classifier can be constructed by aggregating the knowledge from local classifiers. It is well suited to small scale distributed classification, but does not involve dynamism of P2P networks. It also suffers from heavy communication overhead.

Research based on Distributed Plurality Protocol in dynamic P2P networks for combining results from local classifiers is described in [32]. It describes an ensemble paradigm for P2P classification in which each peer builds its local classifiers by the learning algorithm of pasting bites.

Another research towards P2P classification is a decision tree induction algorithm which is used for data classification which is described in [33]. Building decision trees is a very challenging task in a P2P network due to its asynchronous nature, huge number of data sources, and the dynamic nature of the data. In [33] presents a distributed and asynchronous algorithm *PeDiT* which induces a decision tree over a P2P network in which every peer has a set of learning examples. This algorithm is motivated by ID3 and C4.5, which aims to select at every node, the attribute that maximize a gain function. Then, *PeDiT* aims to split the node, and the learning examples associated with it, into two new leaf nodes and this process continues recursively. A stopping rule is used for termination. This algorithm seems to be efficient with low communication overhead, accurate, asynchronous, and adapts smoothly to changes in the data and the network.

In [34], a communication-efficient SVM cascade approach to perform distributed classification in P2P networks is presented. This approach exploits the characteristics of DHT-based lookup protocols in order to perform classification and found to be robust to multiple parameters like number of peers in the network, imbalanced data sizes and failure rate of peers . Communication cost is low and accuracy is more for this approach.

### 7.2.3 P2P Clustering

Researches on P2P clustering introduces an exact local algorithm for monitoring k-means clustering described in [36].The k-means monitoring algorithm monitors the distribution of centroids across peers and thus make the k-means process aware of when to update the clusters. The K-means monitoring algorithm consists of two phases - First phase is monitoring the data distribution which is carried out by an exact algorithm, Second phase is centroid computation by a centralization approach. However this algorithm cannot be used to solve distributed clustering.

A modified k-means algorithm based on probe-and-echo mechanism is described in [40]. In this algorithm centers are broadcasted to all peers in the network using probe –and-echo mechanism. Synchronization is needed in each iteration between all the peers and that result in heavy traffic and congestion in the network. A non-locally synchronized k-means algorithm which uses random sampling is described in [41], in which message load is reduced.

Two approximate local algorithms LSP2P and USP2P for P2P K-means clustering are described in [37] and [42]. Local synchronization based P2P K-means distribute the centroids using gossiping. The centers at each peer are updated making use of the information received from their immediate neighbors. This algorithm produces highly accurate clustering results but no analytical guarantees on this clustering accuracy is provided. So a second algorithm Uniform Sampling based P2P K-means is developed. Probabilistic guarantees are provided through sampling.USP2P assumes the network as static and found to achieve high accuracy. Both these algorithms are based on the assumption that data distribution among the peers is uniform. So it may not work well for large size networks. Also since text collections in P2P networks may not be uniformly distributed, these algorithms do not suit for text collection.

In [39] describes a frequent term based text clustering algorithm for P2P networks. Initially, the maximal frequent term sets of all documents are found. For each maximal frequent term set, the local initial centers are computed. An approximate global initial cluster centers is got via communication of each node with its direct neighbors. Finally local documents are clustered based on the initial cluster centers. Communication overhead is very less and cluster quality will not decrease as size of the network grows.

Modifications to approximate K-means algorithm is described in [38]. An efficient distributed algorithm that constructs clustering over large database based on HDP2P architecture is proposed. Communication is only among immediate neighbors and synchronization is local. The algorithm is found to achieve effective clustering result and low communication cost.

## 8. CONCLUSION AND FUTURE WORK

P2P data mining is recently emerged from distributed data mining which deals with data analysis in environments with distributed data, computing nodes, and users. This focuses on a literature survey on distributed data mining in P2P networks. First of all we discussed the centralized data mining approach and the need for distributed data mining. In the continuing sections we saw the distributed data mining approach and a classification of various DDM architectures. Then P2P data mining, the motivation behind P2P, issues and challenges of P2P data mining are discussed.

Data mining in P2P environment need to consider a lot of features like scalability, fault tolerance, decentralization, communication efficiency, anytimeness and asynchronism. Only few P2P data mining algorithms are emerged so far and integrating them with real P2P applications involve many challenges. This survey reveals that even though P2P data mining algorithms addresses some of the above mentioned features one or other, none of them cover all the required features. Research towards developing such algorithms need to be considered. More works that incorporates optimization techniques like ant colony optimization with P2P data mining are to be considered. Mobile P2P is a recently emerging area and future works on mobile P2P based data mining are also encouraged.

[45]    www.arpapress.com/Volumes/Vol4Issue4/IJRRAS_4_4_06.pdf

[46]    Napster: http://www.napster.com/

[47]    Eng Keong Lua, Jon Crowcroft, Marcelo Pias, Ravi Sharma and Steven Lim "A Survey and Comparison of Peer-to-Peer Overlay Network Schemes" IEEE Communications Surveys and Tutorials - COMSUR , vol. 7, no. 1-4, pp. 72-93, 2005

[48]    S. Ratnasamy, P. Francis, M. Handley, R. Karp, and S. Shenker, "A scalable content addressable network," in Processings of the ACM SIGCOMM, 2001, pp. 161–172.

[49]    B. Y. Zhao, L. Huang, J. Stribling, S. C. Rhea, A. D. Joseph, and J. D.Kubiatowicz, "Tapestry: A resilient global-scale overlay for service deployment," IEEE Journal on Selected Areas in Communications,vol. 22, no. 1, pp. 41–53, January 2004.

[50]    Stoica, R. Morris, D. Karger, M. F. Kaashoek, and H. Balakrishnan, "Chord: A scalable peer-to-peer lookup protocol for internet applications," IEEE/ACM Transactions on Networking, vol. 11, no. 1, pp. 17– 32, 2003.

[51]    D. Malkhi, M. Naor, and D. Ratajczak, "Viceroy: a scalable and dynamic emulation of the butterfly," in Processings of the ACM *PODC'02*, Monterey, CA, USA, July 2002, pp. 183–192.

[52]    Clarke, O. Sandberg, B. Wiley, and T. W. Hong. (1999) Freenet: A distributed anonymous information storage and retrieval system. Freenet White . [Online]. Available: http://freenetproject.org/ freenet.pdf

[53]    Gnutella development forum, the gnutella v0.6 protocol.[Online]. Available: http://groups.yahoo.com/group/the gdf/files/

[54]    Fasttrack Peer-to-Peer Technology company. [Online]. Available: http://www.fasttrack.nu/Kazaa media desktop. [Online]. Available: http://www.kazaa.com/
Survey on Distributed Data Mining in P2P Networks                                                    23